\documentclass[12pt]{article}

\usepackage{amsfonts}
\usepackage{amsmath}

\newcommand{\uiaddress}
{{\small\it Department of Physics, University of Illinois, Urbana, IL 61801}}
\newcommand{\email}[1]{\thanks{e-mail: \tt#1}}
\newcommand{\preprint}
{\begin{flushright}\begin{small}
ILL-(TH)-00-08\\ hep-th/0009209\\ 
\end{small}\end{flushright}       
}

\newcommand{\pa}{\partial}
\newcommand{\tr}{{\rm tr}}
\newcommand{\comment}[1]{}

\newcommand{\pasl}{\pa\kern-.55em /}

\newcommand{\ksl}{k\kern-.55em /}

\DeclareFixedFont{\xiiss}{OT1}{cmss}{m}{n}{12}
\DeclareFixedFont{\ixss}{OT1}{cmss}{m}{n}{9}
\DeclareFixedFont{\cmrnine}{OT1}{cmr}{m}{n}{9}
\newcommand{\field}[1]{\mathbb{#1}}
\newcommand{\BC}{{\field C}}

\newcommand{\BZ}{{\field Z}}

\newcommand{\CCs}{\hbox{\ixss C\kern-.4emI}}
\newcommand{\ZZs}{\hbox{\ixss Z\kern-.4emZ}}

\newcommand{\CA}{{\cal A}}

\newcommand{\CM}{{\cal M}}
\newcommand{\CZ}{{\cal Z}}

\newcommand{\ZA}{\CZ\CA}

\newcommand{\CP}{{\BC\field P}}
\newcommand{\CY}{CY}

\newcommand{\diag}{\hbox{diag}}
\newcommand{\cst}{$C^*$ algebra}

\begin{document}

\begin{titlepage}
        \title{
        \preprint\vspace{1.5cm}
		Non-Commutative Calabi-Yau Manifolds}
		\author{
        	David Berenstein,\email{berenste@pobox.hep.uiuc.edu}
		\\
				and\\     
		Robert G. Leigh\email{rgleigh@uiuc.edu}\\
	\uiaddress
        \\
		}
\maketitle

\begin{abstract}
We discuss aspects of the algebraic geometry of compact non-commutative 
Calabi-Yau manifolds. In this setting, it is appropriate to consider
local holomorphic algebras which can be glued together into a compact
Calabi-Yau algebra.
We consider two examples: a toroidal orbifold $T^6/\BZ_2\times\BZ_2$,
and an orbifold of the quintic in $\CP_4$, each with discrete torsion.
The non-commutative geometry tools are enough to describe various properties
of the orbifolds. First, one describes correctly the fractionation
of  branes at singularities. Secondly, for the first example we show that
one can recover explicitly a large slice of the moduli space of complex structures
which deform the orbifold. For this example we also show that we get
 the correct counting of complex structure deformations at the 
 orbifold point by using traces of non-commutative differential forms
 (cyclic homology).
\end{abstract}
\end{titlepage}

\section{Introduction}

Studying D-branes by the boundary state formalism is in general
difficult. In particular one needs an exact closed string conformal
field theory (for reviews, see \cite{gaber,Drev}). Although these are
powerful techniques, once one deforms the theory, the boundary states
may be difficult to follow through the deformation. One can also study
D-branes on smooth manifolds in the large volume limit by means of
algebraic geometry, but there are known examples  of string compactifications
where singularities
cannot be resolved completely (for example, orbifolds with discrete
torsion\cite{VW}). In this case a commutative algebraic geometric
resolution gives the wrong counting of deformations. This signals that
commutative algebraic geometry methods can be deficient in extracting the 
proper geometry for a string background which is singular.

In previous work \cite{BJL1, BJL2} we have shown that following marginal
and relevant supersymmetric deformations of a conformal field theory
associated to the near horizon geometry of a collection of $D3$ branes
gives rise to non-commutativity in the moduli space of vacua. These
field theories are related to orbifolds with discrete torsion, and we
began work on a systematic treatment of non-commutativity from a
holomorphic algebro-geometric point of view. The objective of the
present letter is to show several examples of {\it compact}
non-commutative geometries. We will show that this provides the proper
set of tools for describing the aforementioned singularities. 
The non-commutativity is
inherent and may not be removed by deformations. The process of
compactification is a prescription for glueing together local
non-commutative algebras.

In Section \ref{sec:review} we review the formulation of non-commutative
algebraic geometry used in \cite{BJL1}. In Sections
\ref{sec:orb1},\ref{sec:orb2} we give two examples; these are orbifolds
with discrete torsion of (a) a 6-torus and (b) the quintic in $\CP_4$.
The first example is treated in more detail as we can calculate exactly
the global deformations of the algebra, although due to some simplifying
choices only a certain class of these are studied directly here. These
are seen to be associated to deformations of the complex structure of
the Calabi-Yau space (namely, the holomorphic constraints on the
non-commutative algebra), and we find agreement with the results of Vafa
and Witten \cite{VW} in the sense that all of the deformations are
allowed by closed string considerations. At the orbifold point, all
local deformations are accounted for by traces of differential forms
({\it i.e.,} by cyclic homology).

\section{An approach to non-commutative algebraic geometry}\label{sec:review}

In this section we will give a condensed version of the geometrical
ideas described in \cite{BJL1}. The idea is to begin with a
(non-commutative) algebra $\CA$ over $\BC$ that is taken to be the
(local) algebra of holomorphic functions on a (non-commutative)
algebraic space, which we label $\CM_\CA$. The center of the algebra
$\CA$, labelled $\ZA$, is the subalgebra consisting of elements of $\CA$
which commute with $\CA$.

The center is a commutative algebra, and as such describes an affine
space $\CM_{\ZA}$. The natural inclusion of the center into the algebra
is to be taken as pullback of functions, so we have a map
\begin{equation}
\CM_\CA \to \CM_{\ZA}
\end{equation}
This local  algebra as given is not a \cst, as there is no
complex conjugation, only holomorphic variables. We can make a 
\cst\ if we introduce the complex conjugate variables to $\CA$, and
make non-holomorphic combinations which are bounded. Thus in principle
we can construct an associated system of local \cst s which
can be embedded into a larger \cst\ $\tilde \CA^*$. An example is the usual 
quantum torus: there is the algebraic relation $UV=qVU$, and additional
unitarity conditions $U^{-1}=U^\dagger$ and $V^{-1}=V^\dagger$.

A central idea is that there are two geometries: a commutative geometry on which
closed strings propagate, and a non-commutative version for open strings \cite{SW}.
Here, we identify the commutative space with the center ${\ZA}$ of the algebra.
The center of the algebra describes
locally a Calabi-Yau manifold with  singularities which is
also taken to be large so that a semiclassical string analysis is valid.
As there are singularities, the commutative geometry alone does not describe
how these are resolved, and it is here that the non-commutative geometry
gives the extra information needed to provide the boundary conditions for 
twisted states. Thus, the twisted states will be sensitive to the
non-commutative geometry at the singularities. 
We will give a description of the massless
closed string states tha count complex structure 
deformations directly with  non-commutative geometry tools, as such 
we will not compute vertex operators for twisted states directly.
We will show that our defintion gives the right counting of states, 
but the connection between the two possible 
descriptions is not clear.

Point-like $D$-branes on the geometry associated to $\CA$ are 
constructed algebraically as irreducible representations of
$\CA$, where two representations related by conjugation by
$GL(n,\BC)$ are identified. 
We will require that these irreducible representations are finite
dimensional and thus given by matrices, and moreover that the dimension
associated to the center of the algebra is the compactification dimension.

On each of these representations Schur's lemma forces the elements of
the center to be proportional to the identity, and as such the
representation defines a point in the commutative geometry of the
center. Thus we can say where on the Calabi-Yau space the point like
D-brane is located.

It is important that the point-like D-branes can probe all of the
geometry of the Calabi-Yau space. Algebras which satisfy these
conditions are referred to as {\it semi-classical}. The quantum torus,
at rational values of $B$, is an example of a semi-classical \cst. In
this case, $U^{\pm n}$ and $V^{\pm n}$ generate the center of the
algebra, where $q^n=1$.

In the examples  studied in \cite{BJL1}, these requirements
seem to  imply that the algebra $\CA$ is finitely generated
as an algebra over $\ZA$. As such, there is an upper bound on the dimension
of the irreducible representations of $\CA$, and the representations
that satisfy this upper bound are taken
to describe a point-like brane in the bulk. 
On taking limits bulk representations can become reducible and this 
phenomenon is interpreted as brane fractionation at singularities. 

%
More generally, we can expect to build holomorphic D-branes as coherent
sheaves over the ring $\CA$. As the center acts on $\CA$, these also
have an interpretation as  coherent sheaves over the commutative
structure, and one can ask if given a coherent sheaf over the center
whether it has an action of $\CA$ which makes it a coherent sheaf on the
non-commutative space. It may be the case that an obstruction to such a
lifting occurs. This would be interpreted as a breaking of the brane in
the non-commutative geometry, and as such it does not exist. This is also
equivalent to the anomaly in brane charge calculated by Witten
for orbifolds with discrete torsion \cite{Wbar,FW}.


Finally, given these algebras one would like to construct closed string
states. These are conjectured to be associated with single-trace
operators including differential forms. This is motivated by ideas in
the AdS/CFT correspondence \cite{M,W,GKP}. The trace is necessary to
render the operator gauge invariant, and differential forms arise in
thinking in terms of vertex operators for the closed strings. 
This is the construction we are looking for, where there is a 
description of closed string states directly from the non-commutative algebra
point of view, which might be generalized to situations where there are
no other constructions available for the closed string states.
We define
the support of a closed string state as the set of points of the
non-commutative geometry where the trace of the state does not vanish.
We will see later in this paper, that this conjecture identifies closed
strings with cyclic homology and that moreover we get the
 right counting of states.

Note that most string states on a space are not holomorphic, thus one
needs to introduce a \cst\ by adding the complex conjugates of the
variables and finding the complete set of commutation relations. One
must do this in such a way that the holomorphic center is in the center
of the full algebra, and that the representation theory of the
holomorphic local algebra gives the representation theory of the full
\cst\ at the same point (perhaps with some subtleties at singularities).
In a field theory context, this second part amounts to solving the
$D$-terms of the theory.

\section{Orbifold of the torus}\label{sec:orb1}

Consider the orbifold  $T^6/\BZ_2\times\BZ_2$ with discrete torsion. The
non-compact version has been studied extensively.\cite{D,DF,BL} Here, we
will use the compact orbifold as a target space for D-branes and show
that the orbifold can be regarded as a non-commutative Calabi-Yau
compact space.

The first thing we need is a model for the orbifold, so we choose to
represent $T^6$ as the product of three elliptic curves, each given in
Weierstrass form
\begin{equation}\label{eq:weier}
y_i^2 = x_i(x_i-1)(x_i- a_i)
\end{equation}
with a point added at infinity for $i=1,2,3$. The $\BZ_2\times \BZ_2$
will act by $y_i\to \pm y_i$ and $x_i\to x_i$ so that $y_1y_2y_3$ is
fixed under the orbifold action. This is necessary in order to satisfy
the \CY\ condition on the quotient space. The four fixed points of the
orbifold at each torus are located at $y_i=0$ and at the point at
infinity. This point at infinity can be  brought to a finite point by a
change of variables
\begin{eqnarray}\label{eq:chavar}
y_i \to y_i' &=& \frac{y_i}{x_i^2}\\
x_i \to x_i' &=& \frac 1 {x_i}\nonumber
\end{eqnarray}
With these two patches, we can cover each elliptic curve of the product.

Now we would like to introduce discrete torsion. It is necessary that
around each of the singularities the local algebra of holomorphic
functions agree with the non-compact case studied in \cite{D,DF,BL}.
This may be accomplished by choosing the $y_i$ to be anti-commuting
variables
\begin{equation}\label{eq:comm1}
y_i y_j = - y_j y_i \quad \hbox{for } i\neq j
\end{equation}
and the $x_i$ to be in the center of the algebra, appropriate to $\BZ_2$
discrete torsion.

From this starting point, we find that $w= y_1y_2y_3$ is in the center
of the algebra. Thus the invariant variables of the orbifold are exactly
the same variables which belong to the center of the local algebra, and
the center of the algebra reproduces the orbifold space as a commutative
algebra. Now consider transforming to another patch by
(\ref{eq:chavar}). Note that since $x_i$ is in the center of the
algebra, there is no ordering ambiguity, and the structure of
(\ref{eq:comm1}) does not change.

To find the points of the compact non-commutative geometry, we consider
the representation theory of the non-commutative holomorphic algebra
locally. At a generic (bulk) point, there is one irreducible
representation of the non-commutative algebra for each commutative point
in the quotient space. Indeed, the following represents the algebra with
discrete torsion
\begin{equation}\label{eq:param}
y_i=b_i \sigma_i
\end{equation}
where the $b_i$ are complex scalars and the $\sigma_i$ are the Pauli
matrices. There is a multiplicity of inequivalent representations here,
one for each set of roots $x_i$ of the Weierstrass forms.

At the fixed planes, this representation becomes reducible as two out of
the three $y_i$ act by zero. Thus we get two distinct non-commutative
points, as there are two different irreducible representations
corresponding to the two eigenvalues of the non-zero $y_k$.

By taking this limit of the representation, we see that the branes
fractionate on reaching the singularity. The non-commutative points of
the singular planes are then seen to be a double cover of the
commutative singular plane, which is a $\CP^1$. The double cover is
branched around the four points $x_k=0,1,a_k,\infty$ and hence the
non-commutative points form an elliptic curve of the form
(\ref{eq:weier}). Around each of these four points there is a $\BZ_2$
monodromy of the representations, which is characteristic of the local
singularity as measuring the effect of discrete torsion \cite{BL}.

\subsection{Global Deformations of the orbifold}

If we are to identify this non-commutative geometry with the orbifold
with discrete torsion, we should in principle be able to reproduce the
deformations of the orbifold. These were calculated in \cite{VW} using
closed string methods. With
this in mind, we must understand how to modify the algebraic
relations so that we get a new geometry which describes this
deformation. These correspond to global deformations; 
in subsection \ref{subsec:infdef}, we will then consider infinitesimal 
deformations of the orbifold.

We proceed by studying the possible deformations of (\ref{eq:comm1})
within a local coordinate patch. The deformation must be of the form
\begin{equation}\label{eq:deformation}
y_i y_j+y_j y_i = c(\CZ\CA)
\end{equation}
where on the right-hand side we have a function of the center of the
local algebra. This deformation breaks quantum symmetries, so as shown
in Ref. \cite{BL}, they are associated with vevs of twisted string
states.

As remarked in \cite{VW}, the generic deformation of the Calabi-Yau
manifold will be a double cover of $\CP^1\times \CP^1\times \CP^1$.
Thus, we should have three variables $x_1,x_2,x_3$ in the center of the
deformed algebra. In the following, to simplify matters, we will assume
that the $y_i$ are given by their Weierstrass form. This restricts the
deformations to a certain class which is not generic; there are other
deformations which are not in this class that we will not discuss here (but
see Section \ref{subsec:infdef}).

As we are doing algebraic geometry, it must be the case that the right
hand side of (\ref{eq:deformation}) is polynomial and free of poles in
each patch. Specifically, (\ref{eq:deformation}) may only take the form
\begin{equation}
y_1 y_2 + y_2 y_1 = 2 P_{12}(x_1,x_2,x_3)
\end{equation}
where $P_{12}$ is a polynomial. Under the change of variables (\ref{eq:chavar}) on
$x_1,y_1$, $P_{12}$ transforms as
\begin{equation}
P_{12}(x_1,x_2,x_3)\to x_1^2 P_{12}(1/x_1,x_2,x_3)
\end{equation}
Thus, as $P_{12}$ should transform into a polynomial, 
$P_{12}$ must be of degree at most two in $x_1$. Similarly, it is at most
degree two in $x_2$ and independent of $x_3$. A similar result
holds for the other two commutation relations, so we get two more
polynomials $P_{13}(x_1,x_3), P_{23}(x_2,x_3)$. Each of these
polynomials has nine parameters.

Now, one sees that $y_1y_2y_3$ does not belong to the center of the
local algebra anymore, and needs to be modified. To see what this
modification is, we compute
\begin{eqnarray}
[y_1y_2y_3, y_1] &=& y_1 y_2 (y_3 y_1+y_1y_3)- y_1y_2 y_1 y_3 - y_1 y_1 y_2 y_3\\
&=& 2 y_1 y_2 P_{13} - 2 y_1 y_3 P_{12}\\
&=& [y_1,y_2] P_{13} - [y_1,y_3] P_{12}
\end{eqnarray}
Similar calculations with $y_2, y_3$ are sufficient to see that 
\begin{equation}
w = y_1 y_2 y_3 +P_{13} y_2 - P_{23}y_1 -P_{12}y_3
\end{equation}
belongs to the center of the local algebra.

Now it is simple to see that $w^2$ is given by a polynomial of degree
$4$ in $x_1,x_2,x_3$. When we do a change of variables to another
coordinate patch as in (\ref{eq:chavar}), then $w$ transforms simply
$w\to \frac{w}{x_i^2}$ and thus correctly describes a double cover of
$\CP^1\times \CP^1\times \CP^1$.

The total number of deformations that we account for in this way is
$30$, nine from each polynomial $P_{ij}$, plus the three $a_i$ that
describe the $y_i$ as a double cover of the $x_i$. The total number of
deformations is actually $51$ \cite{VW} and so we are missing a number of
complex structure deformations.

Indeed, for each pair $P_{ij}$ one should be able to count $16= 4\times
4$ deformations instead of $9=3\times 3$. Notice that we are missing one
deformation for each of the $y_i$; this suggests that in general we
should not be able to interpret the $y_i$ locally as a double cover of
each of the $\CP^1$'s, so our choice of a simple form for the $y_i$ in terms
of the $x_i$ is the reason for the mismatch.

Nevertheless, the deformations that we have considered do resolve all of
the codimension two singularities. Locally the analysis is the same as
in Ref. \cite{D}, so there is a conifold singularity for each of
the codimension three singularities of the orbifold.

\subsection{Local Deformations and Cyclic Homology}\label{subsec:infdef}

Now, let us analyze the possible infinitesimal variations of complex
structure at the orbifold point. For commutative Calabi-Yau manifolds, these are given
by holomorphic sections of $\Omega^{2,1}(\CM)$, the $(2,1)$ forms. 
That is, we need exact forms of the type
\begin{equation}
f d\phi^i\wedge d\phi^j \wedge d \bar\phi^k
\end{equation}
This notion must be modified appropriately for non-commutative geometry.
The deformations of complex structure are related to closed string
states, and from the AdS/CFT correspondence, we expect that these in
turn correspond to single-trace operators. Thus the natural modification
is to include a trace; that is, we consider single trace operators on
the module of differentials. This is precisely the type of objects that appear
in cyclic cohomology \cite{Connes}.

We will study this cohomology directly within the representation
(\ref{eq:param}). Thus, we write
 \begin{equation}
 dy_i = \sigma^i  db_i
 \end{equation}
and we take complex conjugation on the $b_i$ as the $*$-conjugation of the 
$y_i$. That is, $y_i$ commutes with its adjoint $\bar y_i$, and anticommutes
with the adjoints of the other $y_k$.

Now, we are looking for elements of $\Omega^{2,1}(\CM)$ which are
associated with codimension two singularities. A standard way to count
in ordinary orbifolds is to blow up the singularities, and then consider
the cohomology of the smooth manifold. Since the singularities are
resolved by 2-spheres, we may think of a given $(2,1)$-form as giving,
upon integration over the $S^2$, a holomorphic one-form.

Thus in our algebraic setting, to study $\Omega^{2,1}$, we should look
for one-forms such as 
\begin{equation}\label{eq:twistedst}
\tr\ \frac{dy_1}{x_1}=\frac{1}{x_1}\ d\ \tr\ y_1
\end{equation} 
Notice that these are proportional to $\sigma_1$ and hence the trace
vanishes at generic points. As we go to a singularity, say $y_2= y_3=0$,
we should take the trace in distinct irreducible representations of the
algebra. As the branes fractionate at this singularity, $\sigma_1$
decomposes into two one-dimensional irreducible representations and
(\ref{eq:twistedst}) is non-zero. The obvious interpretation is that the
$(2,1)$-form has support only at the singularity -- the above closed
string state couples to fractional branes, and thus should be
interpreted as a twisted sector of the string.

We may repeat this analysis at each of the singular planes; at each, we
get one holomorphic form of the right type which is supported over the
singularity. This construction should be interpreted locally, and not as
giving a single global form on all the planes simultaneously. In this
fashion we get a candidate for each complex deformation in the
(non-commutative) cohomology of differential forms of the space.

Moreover, notice that the support of the closed string state depends on
the branes fractionating at the singularities. Since the branes don't
fractionate further at the codimension three singularities, we don't
expect to see any new (massless) closed string states that are supported
at these singularities. There seems to be no complex structure moduli
supported in codimension three. We conclude that the non-commutative
geometry is capturing all of the string theory data of the orbifold.

\section{Orbifold of the Quintic}\label{sec:orb2}

Now, we consider a second example, a non-commutative version of (an orbifold of) 
the quintic in $\CP^4$\cite{GrP, COGP,CxK}. This example allows us to show how
to glue together patches in a situation less trivial than the toroidal orbifold.

The complex structure moduli space of the quintic has special points with a
$\BZ^3_5$ group of isometries. In this case, we may consider the orbifold
constructed by modding out these symmetries. This orbifold may also include
discrete torsion phases. As we will see, we will actually succeed in taking
only a $\BZ_5\times \BZ_5$ orbifold, but this is sufficient to demonstrate the principles
involved. 

We begin with the quintic described by
\begin{equation}\label{eq:quintic}
{\cal P}(z_j)=z_1^5+z_2^5+z_3^5+z_4^5+z_5^5+\lambda z_1z_2z_3z_4z_5= 0
\end{equation}
where the $z_j$ are homogeneous coordinates on $\CP^4$. The $\BZ_5^3$ action is
generated by phases acting on the $z_j$ as 
$z_j\to \omega^{a_j} z_j$,  with $\omega^5= 1$, and the vectors
\begin{eqnarray}
\vec a &=& (1,-1,0,0,0)\\
&=&(1,0,-1,0,0)\\
&=&(1,0,0,-1,0) 
\end{eqnarray}
consistent with the CY condition $\sum a_i=0\ {\rm mod}\ 5$. 
We have chosen the action such that $z_5$ is invariant. This allows us to 
consider a coordinate patch where $z_5=1$, and this patch is invariant
under the group action. Within this coordinate patch, we
will be interested in the local non-commutative algebra of the other four
variables. Later, we will consider transformations to other patches.

The discrete torsion is classified by \cite{V}
\begin{equation}\label{eq:discrete}
H^2(\BZ_5^3,U(1)) = \BZ_5^3
\end{equation}
so we need three phases to determine the geometry.

The invariant quantities (within the coordinate patch) are given by $z_i^5$ and 
$z_1z_2z_3z_4$. We will require that these remain in the center of the 
non-commutative version of the quotient space.
Experience with standard orbifolds with discrete torsion suggests that we take the variables
to commute up to phases as follows
\begin{eqnarray}\label{eq:ncalgebra5}
z_1 z_2 &=& \alpha z_2 z_1\\
z_1 z_3 &=& \beta \alpha^{-1} z_3 z_1\\
z_1 z_4 &=& \beta^{-1} z_4z_1\\
z_2 z_3 &=& \alpha \gamma z_3 z_2\\
z_2 z_4 &=& \gamma^{-1} z_4 z_2\\
z_3 z_4 &=& \beta \gamma z_4 z_3
\end{eqnarray}
with $\alpha,\beta,\gamma$ being fifth roots of unity. Notice that we
have three fifth roots of unity to choose from, which gives us the same
number of choices as expected from (\ref{eq:discrete}). We will consider
the generic case where all three $\alpha,\beta,\gamma$ are different
from one and we want to find the irreducible representations of this
algebra within the patch $z_5=1$.

%
%
%
%

It is simple to see that there are five-dimensional representations.
Indeed, consider the matrices
\begin{equation}
P = \diag(1,\alpha,\alpha^2,\alpha^3,\alpha^4), \quad Q = \begin{pmatrix} 0 & 0 &0 & 0 &1\\
1 &0 &0 & 0 &0\\
0 & 1 & 0 &0 &0\\
0 & 0 & 1 & 0 & 0\\
0 & 0 &0 &1 &0
\end{pmatrix}
\end{equation} 
$\alpha$ is a fifth root of unity, and so there are actually five distinct 5-dimensional
representations here.
In terms of the matrices $P,Q$, we can identify $ z_1 = b_1 P$ and $z_2 = b_2 Q$, where $b_j$
are arbitrary complex numbers. Similarly, we may write $z_3 =b_3 P^m Q^n$ and
$z_4 = b_4 P^{-m-1} Q^{-n-1}$, where $m,n$ are determined by the phases $\beta,\gamma$
(i.e., $\beta=\alpha^{n+1}$ and $\gamma=\alpha^{m+1}$).
These choices guarantee that $z_1 z_2 z_3 z_4 z_5$ is in the center of the
algebra.

The defining relation (\ref{eq:quintic}) requires that
\begin{equation}\label{eq:bvari}
\sum_{i=1}^4 b_i^ 5 +\lambda\prod_{i=1}^4 b_i= -1
\end{equation}

As given, the representation becomes reducible only when three out of
the four $z_i$ act by zero, which signals that there are no fractional
branes for the (would be) generic codimension two singularities. Indeed
one can see that this is the case locally, as there are other invariant
variables in $z_1, z_2, z_3$ apart from $z_1^5, z_2^5, z_3^5$, which
signals that we have only done a $\BZ_5$ orbifold locally.

At co-dimension three singularities, we do get the expected number of
fractional branes however, which is five, so our analysis agrees with a
direct analysis local to the singularity \cite{DM}. We also recover the
local quiver algebra by noticing that at least one of the $z_j$ has a
large vev (see eq. (\ref{eq:bvari})), and this field is responsible for
breaking the gauge group from $U(5)$ to $U(1)^5$. Also, ensuring that
the commutation relations with this large field hold forces many of the
matrix terms in the other fields to vanish, thus giving the correct
local data to reconstruct the quiver diagram of the orbifold.

This fact can be traced back to how the discrete torsion lattice is
related to the orbifold group. Indeed, we expect D-brane states to be
given by a formal quiver diagram corresponding to the irreducible
projective representations of the group $\Gamma$ with a given cocyle
\cite{D, G}. No matter what choice we make for the cocyle condition, as
long as it is not trivial, there are  $5$ different inequivalent
irreducible representations of the group $\Gamma$, so we expect to see a
quiver diagram with five nodes instead of one as we have written naively
above. We will explore the construction of the non-commutative algebraic
geometry associated to quiver diagrams in  future papers \cite{BJL3,BJL4}, so
we will leave this issue at this point.

Our choice of phases has not given the full expected result. Instead,
two of the phases select a $\BZ_5^2 \subset \BZ_5^3$ subgroup, and the
third phase determines the discrete torsion phase for the generators of
this subgroup. Thus, the algebra written in (\ref{eq:ncalgebra5}) is
describing a generic $\BZ_5^2$ orbifold of the quintic with discrete
torsion.

In a non-generic case we can choose the $\BZ_5^2$ to have singularities
in codimension two. In this case we choose $z_4$ to be in the center of
the algebra, and we have only the phase given by $\alpha$ to worry
about. There we see the familiar result for orbifolds with discrete
torsion appearing. The branes fractionate in the codimension $2$
singularities, and they have the expected monodromies about the
codimension three singularities \cite{BL}.

Now let us consider coordinate transformations to other patches. 
Indeed, it is not difficult to show that one can still do the 
standard coordinate changes of the quintic; one must  only be 
careful with ordering of variables.
In every coordinate patch the algebra obtained is of the same type as written in
(\ref{eq:ncalgebra5}).

Let us consider the following example, 
\begin{equation}
z_i \to  z^{-1}_1 z_i
\end{equation}
As $z_1^5$ is in the center of the algebra,
we may interpret this as
\begin{equation}
z_i\to \frac1{z_1^5} z_1^4 z_i.
\end{equation}
Thus, denominators may always be taken to lie in the center of the
algebra. The ordering of the non-central part in the numerator is
irrelevant; the difference between any two choices amounts to a phase,
which may be removed by a trivial change of variables.

Algebraically this means that localization takes place in the center.
That is, changes of variables can be first done in the center and then
lifted to the non-commutative algebra, which fits nicely with the
structure proposed in \cite{BJL1} where the center takes a prominent
role in the non-commutative algebraic geometry of matrices.

\section{Outlook}

We have shown in this paper that non-commutative algebraic geometric
tools are sufficient to describe various compact orbifolds with discrete
torsion and their deformations.

The prescription involves gluing various non-commutative holomorphic
algebras to compactify spaces. Then, one can calculate the deformations
of complex structure by using the cyclic homology of the glued algebras,
and obtain exact agreement with the calculations done by Vafa and Witten
\cite{VW} with closed string methods.

The non-commutative geometry thus gives a framework in which to
interpret backgrounds which do not follow the standard rules of
commutative algebraic geometry. In principle, this provides a new set of
tools by which one can construct new backgrounds of string theory. These
do not have to come from global orbifolds with discrete torsion,  and
may be singular when viewed from  a commutative geometry point of view,
while the non-commutative geometry is  under control.

The second point which is important is that topological closed string
states are associated with cyclic homology. Thus, one should be able to
extend this idea beyond the classes of algebras which we have studied.
This perspective might shed some light on aspects of string geometry.

The extension of these ideas to more general orbifolds and backgrounds
is currently under consideration \cite{BJL3, BJL4}.

\noindent {\bf Acknowledgments:} We wish to thank M. Ando and V. Jejjala
for discussions.
 Work supported in part by U.S.
Department of Energy, grant DE-FG02-91ER40677 and 
an Outstanding Junior Investigator Award. 

\providecommand{\href}[2]{#2}\begingroup\raggedright\endgroup

\end{document}